\documentclass[twocolumn]{aastex61}
\usepackage{graphicx,xcolor}

\submitjournal{ApJ}

\shorttitle{Torsional Oscillations contribute to the Waldmeier effect}

\begin{document}

\title{Torsional Oscillations in the Sun's Rotation contribute to the Waldmeier-effect in Solar Cycles}

\author{Sushant S. Mahajan}
\affil{Department of Physics, Indian Institute of Technology (Banaras Hindu
University), Varanasi 221005, India}
\affil{Center for Excellence in Space Sciences India and Department of Physical Sciences, Indian Institute of
Science Education and Research, Kolkata, Mohanpur 741246, India}
\affil{Currently at: Department of Physics \& Astronomy, Georgia State University, Atlanta 30303, GA, USA}

\author{Dibyendu Nandy}
\affil{Center for Excellence in Space Sciences India and Department of Physical Sciences, Indian Institute of
Science Education and Research, Kolkata, Mohanpur 741246, India}

\author{H.M. Antia}
\affil{Tata Institute of Fundamental Research, Homi Bhabha Road, Mumbai
400005, India}

\author{B.N. Dwivedi}
\affil{Department of Physics, Indian Institute of Technology (Banaras Hindu
University), Varanasi 221005, India}

\correspondingauthor{Sushant S. Mahajan}
\email{mahajan@astro.gsu.edu}

\begin{abstract}

Temporal variations in the Sun's internal velocity field with a periodicity of
about 11 years have been observed over the last four decades. The period of these
torsional oscillations and their latitudinal propagation roughly coincides with
the period and equatorward propagation of sunspots which originate from a
magnetohydrodynamic dynamo mechanism operating in the Sun's interior. While the
solar differential rotation plays an important role in this dynamo mechanism by
inducting the toroidal component of magnetic field, the impact of torsional
oscillations on the dynamo mechanism -- and hence the solar cycle -- is not well
understood. Here, we include the observed torsional oscillations into a flux
transport dynamo model of the solar cycle to investigate their effect. We find that
the overall amplitude of the solar cycle does not change significantly on
inclusion of torsional oscillations. However, all the characteristics of the
Waldmeier effect in the sunspot cycle are qualitatively reproduced by varying
only the amplitude of torsional oscillations. The Waldmeier effect, first noted
in 1935, includes the important characteristic that the amplitude of sunspot
cycles is anti-correlated to their rise time; cycles with high initial rise rate
tend to be stronger. This has implications for solar cycle predictions. Our
results suggest that the Waldmeier effect could be a plausible outcome of cycle to cycle modulation of torsional oscillations and provides a physical basis for sunspot cycle forecasts based on torsional oscillation observations. We also provide a theoretical explanation based on the magnetic induction equation thereby connecting two apparently disparate phenomena.

\end{abstract}

\keywords{ magnetic fields; Sun: activity; Sun: dynamo; Sun: interior}

\section{Introduction} \label{sec:intro}

\begin{figure}[h!]
\begin{center}
  \includegraphics[width=9 cm]{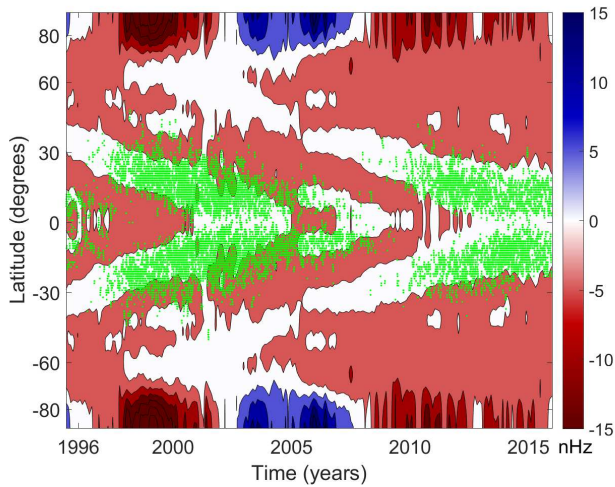}\\
  \caption{The observed plot of torsional oscillations in the solar rotation (in nHz) on the solar
surface from GONG measurements is depicted here. The butterfly diagram of locations of
sunspots (denoted by green dots) for cycles 23-24 obtained from the Royal Greenwich Observatory
\citep{RGO} is over-plotted. The similar spatio-temporal evolution and periodicity of the sunspot cycle and torsional oscillations indicate a link between the two.}\label{tor+bfly}
  \end{center}
\end{figure}

Helioseismic observations over the past four decades have made it possible to
study the Sun's internal velocity fields. The Sun's azimuthal rotation rate,
which is responsible for the generation of sunspot producing toroidal magnetic field, was believed to be
constant in time until \citet{1980ApJ...239L..33H} found torsional waves
propagating towards the equator on the solar surface from the rotation rate
measurements at Wilcox solar observatory. About a decade later, the addition of
Helioseismology to the bag of tools used to study the Sun provided the
opportunity to precisely measure the plasma flows inside the Sun. Helioseismic
measurements of the azimuthal rotation rate using data from Big Bear
Solar Observatory (BBSO; \citet{1993Sci...260.1778W}), Michelson Doppler Imager
(MDI; \citet{1998ApJ...505..390S}) and Global Oscillations Network Group (GONG;
\citet{2000SoPh..192..427H}) confirmed the existence of torsional waves even in the solar interior. These
waves, now known as torsional oscillations (see Fig. \ref{tor+bfly}), appear
as
bands of faster and slower than average rotation propagating towards the
equator in the low-latitude region below 60 degrees and towards the poles in
the high-latitude region above 60 degrees \citep{2001ApJ...559L..67A}.

The similarity in the spatio-temporal evolution of the sunspot belt and
torsional oscillation pattern \citep{1982SoPh...75..161L} has led many
researchers to consider direct Lorentz feedback of the magnetic field on plasma
flows, (magnetically mediated) geostrophic flows or other indirect energy
transfer mechanisms as plausible causes of torsional oscillations
\citep{2000A&A...360L..21C,2003SoPh..213....1S,2006ApJ...647..662R,2013SoPh..282..335B,2016ApJ...828L...3G}. Models based on these theories have been quite
successful in producing torsional oscillation patterns similar to the patterns
observed.

Past studies have primarily focussed on exploring theories for the origin of torsional oscillations with the aim of reproducing the observed pattern of torsional oscillations based on feedback mechanisms on the solar plasma. But the
generation of torsional oscillations is only half of the story. In this paper we
explore the other half, that is once the torsional oscillations
are taken into account how do they alter the magnetic field induction process.
We achieve this by incorporating the helioseismic measurements of torsional
oscillations from GONG observations into a flux transport dynamo model, which
explores the effect of plasma flows in the solar interior on the magnetic field induction process.  The premise of our numerical
experimentation is based on two considerations. One that the magnetic feedback
on the plasma flows is weak -- which is observed and established theoretically
\citep{2006ApJ...647..662R}; this allows for the overall flux transport
principle to be effective. Second, variations in the amplitude of torsional
oscillation may impact the nature of the sunspot cycle through
changes to the Sun's internal differential rotation profile.

In section \ref{data}, we give a brief description of the helioseismic data from
GONG that has been used to carry out
this study. Then we go on to describe the flux transport dynamo model we have
used in this study in section \ref{model}, and
how the torsional oscillations have been introduced in this model in section
\ref{torinclusion}. We discuss the results obtained in section \ref{results} and provide a theoretical explanation for our findings in section \ref{explanation}. The implications of the results are discussed in section \ref{conclusion}.

\section{Methods}\label{methods}

\subsection{Helioseismic data on torsional oscillations}\label{data}

We use the helioseismic data from the Global Oscillations Network Group (GONG)
project (Hill et al.~1996) covering the period from 7 May 1995 to 9 September
2012.
This data consists of 174 sets each covering a period of 108 days with a shift
of 36 days between
successive data sets. This period covers the entire solar cycle 23 and the
rising phase of cycle 24.
Each data set consists of frequencies and splitting
coefficients for p-modes up to a degree $l=150$ and for each set we performed
a 2d Regularised Least Squares (RLS) inversion \citep{1998MNRAS.298..543A} to calculate
the rotation
rate as a function of radius and latitude. This gives
us the temporal variation in the solar rotation rate. To isolate the temporally
varying part associated with torsional oscillations observed at the solar
surface we subtract the mean rotation rate at each radius and latitude from
the value at each time to get the residual:
\begin{equation}
\delta\Omega(r,\theta,t)=\Omega(r,\theta,t)-\langle\Omega(r,\theta,t)\rangle
\label{eq:zonal}
\end{equation}
where the angular brackets denote the temporal average over the period of
solar cycle 23 which was estimated to be 11.7 years \citep{2010ApJ...720..494A} long.

\subsection{Flux transport dynamo model of the solar cycle}\label{model}

The solar magnetic cycle originates via a dynamo mechanism which recycles the
toroidal and poloidal components of the solar magnetic field relying on solar plasma flows. For this study we utilize a flux transport
dynamo model which has been well studied in different contexts
\citep{2002Sci...296.1671N,2004A&A...427.1019C,2008ApJ...673..544Y, 2014A&A...563A..18P}. This model
solves for the magnetic induction equation (in the solar convection zone) given
by:

\begin{equation}\label{ind}
\frac{\partial B}{\partial t}=\nabla \times (v\times B - \eta \nabla\times B).
\end{equation}
Assuming axisymmetry, one can split the magnetic field into the toroidal and
poloidal components as in equation \ref{split}, and the velocity as in equation
\ref{vel}
\begin{eqnarray}\label{split}
B&=&Be_{\phi}+\nabla\times(Ae_{\phi}),\\
\label{vel}
v&=&v_{p}+r\sin\theta\Omega e_{\phi}.
\end{eqnarray}

Plugging equations \ref{split} and \ref{vel} into the induction equation gives
us the equations for the evolution of the toroidal and poloidal field
respectively.
\begin{equation} \label{pol}
\frac{\partial A}{\partial
t}+\frac{1}{s}(v.\nabla)(sA)=\eta_{p}(\nabla^{2}-\frac{1}{s^{2}})A+\alpha B,
\end{equation}
\begin{eqnarray} \label{tor}
\frac{\partial B}{\partial t}+\frac{1}{r}\bigg(\frac{\partial}{\partial r}(r v_{r} B)
+\frac{\partial}{\partial \theta}(v_{\theta} B)\bigg)
=\eta_{t}\bigg(\nabla^{2}-\frac{1}{s^{2}}\bigg)B \nonumber \\
+s(B_{p}.\nabla)\Omega+\frac{1}{r}\frac{d \eta_{t}}{dr}\frac{\partial(rB)}{\partial
r}
\end{eqnarray}
where $A$ is the magnetic vector potential for the poloidal field
($B_{p}$), $B$ is the toroidal field, $v$ is the meridional flow velocity,
$\Omega$ is the azimuthal rotation velocity, $\eta$ is the turbulent magnetic
diffusivity and $s=r\sin\theta$. These two equations govern the two main mechanisms of the solar dynamo:
\vspace{3mm}\par
1. The $\Omega$-effect: The second term on the right hand side of equation
\ref{tor} suggests that the poloidal field ($B_{p}$) would be sheared by the
gradient of the azimuthal rotation velocity ($\Omega$) and get wrapped around
the Sun to produce magnetic field in the toroidal direction. This effect, known
as the $\Omega$-effect, was first proposed by
\citet{1955ApJ...122..293P}.
\vspace{3mm}\par
2. The Poloidal Source: The second term on the right hand side of equation
\ref{pol} acts as a source term for the poloidal field. In this model it is
assumed that the poloidal source arises
completely from the Babcock-Leighton mechanism proposed by
\citet{1961ApJ...133..572B}
and \citet{1969ApJ...156....1L}. They proposed that magnetically buoyant
toroidal flux tubes are tilted due to the Coriolis force (and erupt as tilted
bipolar sunspot pairs) such that they produce a
net poloidal component of magnetic field. This results in the conversion of the
toroidal field back into the poloidal field through near surface flux transport processes.

We use the same diffusivity profile as \citet{2009ApJ...698..461M} which makes
our model a high diffusivity model. The meridional flow velocity profile was
modelled as in \citet{2002Sci...296.1671N}, \citet{2008ApJ...673..544Y} and
\citet{2014A&A...563A..18P} with a small component of the flow penetrating
beneath the base of the solar convection zone. For computational efficiency we
limit our simulation domain to the solar northern hemisphere with a boundary
condition at the equator which imposes dipolar parity of magnetic field. We note
here that the exact nature or values of the parameters above would not impact
the results qualitatively in any case as long as they are within the operational
regime of the flux transport dynamo. Therefore, for ease of comparison, we have
kept the above driving parameters the same as in previous studies. Here we focus
only on the impact of inclusion of torsional oscillations (with a variable
amplitude) on the solar dynamo.

\subsection{Inclusion of Torsional Oscillations in the Flux transport
dynamo}\label{torinclusion}

\begin{figure}[h]
 \centering
 \includegraphics[width=9 cm]{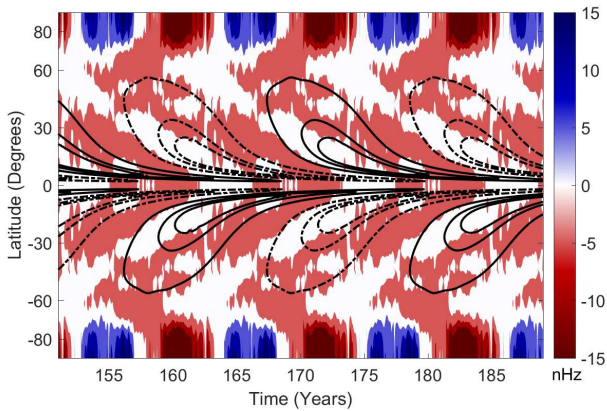}
 \caption{Background: Torsional oscillation profile, obtained from the
repetition of 11.23 years long dataset (GONG) for the northern hemisphere (color axis is the same as in Fig. \ref{tor+bfly}). Foreground: Contour plots for the toroidal magnetic field at the base of the convection zone from flux transport dynamo simulations. Positive and negative magnetic fields are shown with dashed and solid black contours, respectively.}
 \label{period}
\end{figure}

Torsional oscillations are temporal variations in the rate of azimuthal rotation
($\Omega$) in the solar convection zone. Since the amplitude of torsional
oscillations
($20$ nHz) is much smaller than the Sun's average rotation rate (around $400$
nHz), we can include these
oscillations in the flux transport dynamo ($0.55R_{\odot}$ to $1.0R_{\odot}$ and $0$ to $88$ degrees latitude) by augmenting the mean azimuthal
rotation rate ($\Omega_{o}$) by a perturbed quantity ($\delta\Omega$) which
represents
the torsional oscillations.
\begin{equation}\label{omega}
\Omega(r,\theta,t)=\Omega_{o}(r,\theta)+\delta\Omega(r,\theta,t)
\end{equation}
Although the temporal variation in $\Omega$ is small, the dynamo equations
involve the gradient of $\Omega$ and the temporal variations in both radial
and latitudinal gradient of rotation rate can be much larger (Antia et
al.~2008). Hence the effect of temporally varying rotation rate on the solar
dynamo requires careful analysis.

Since both the torsional oscillations and the solar cycle are periodic
phenomena, one needs to handle their phase relationship appropriately, and
ensure that the simulations are initiated with the appropriate relative phase.
The data for torsional oscillations from GONG start from $29^{th}$ June 1995 and
continue till $17^{th}$ July 2015. If we look at the polar field measurements by
Wilcox Solar Observatory
\citep{1995SSRv...72..137H,1978SoPh...58..225S}, we find that the
polar field was maximum in 1995, and was positive in the northern hemisphere. To
reproduce the polar field configuration during 1995, we run the dynamo model
without torsional oscillations (with just the mean rotation rate)
for a few cycles until the solution becomes stable, and then stop it at a
time when the polar field in the northern hemisphere becomes maximum and
positive.
The magnetic field configuration at the end of this simulation is stored and
serves as the initial condition for simulations with torsional oscillations
included in the model.

The observed period of the solar cycle, as well as the period of torsional
oscillations undergoes statistical variations, and is not fixed. Thus, the
period of
torsional oscillations may not always be exactly equal to the period of the
sunspot cycle.
The output of the flux transport dynamo model, however, is a uniformly periodic
cycle. Thus, we have to prepare a torsional oscillation profile that has a
uniform
periodicity so that it can retain its phase relationship with the dynamo output.
This has been done by repeating a segment of the torsional oscillations
data in time. \citet{2010ApJ...720..494A} have shown that the torsional
oscillation pattern had a period of about 11.7
years for the solar cycle 23. With the standard set of parameters the solar
dynamo model used here has a cycle period of 11.23 years. Thus, we have
compressed 11.7 year long torsional
oscillation data starting from 1995 into 11.23 years to match with the cycle
output from the dynamo model. This ensures that the phase relationship between
the solar cycle and the torsional oscillation pattern is maintained as
shown in Fig. \ref{period}. Note that this phase-locking is necessary for utilization of the observed torsional oscillations in a theoretical model and does not impact the sanctity of the simulations in any way.

\section{Results}\label{results}

With the phase and the period of torsional oscillations matched with the solar
cycle, we vary the amplitude of the torsional oscillation to perform a set of
computational simulations as a numerical experiment. Nothing is known about the
variation of the amplitude of torsional oscillations yet, as we have continuous
observations from GONG for only over one and a half solar cycles. Nevertheless,
one expects the amplitude of torsional oscillations to be modulated from one
cycle to another. It is our aim here to study the impact of this varying
torsional oscillation amplitude on the solar cycle. Without making any changes
to the spatio-temporal dependence, we simply change the amplitude of the torsional oscillations by
multiplying the torsional oscillation input to the dynamo model with a scaling factor (sc).

\subsection{Impact of the amplitude of torsional oscillations on the strength of the solar cycle}
\label{sectionamp}

\begin{figure}
  \centering
  \includegraphics[width=8.5 cm]{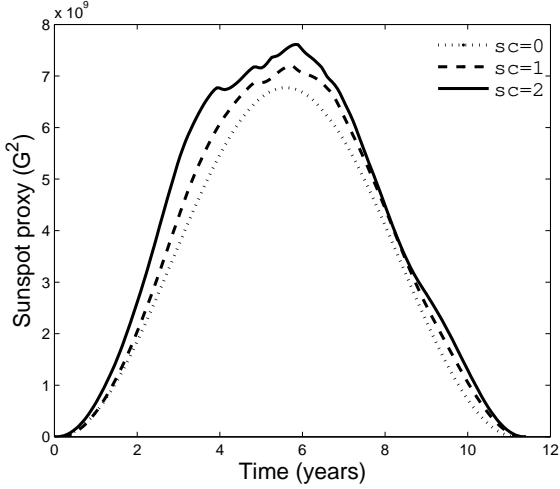}
  \caption{The temporal profile of sunspot proxy ($B^{2}$) for different scaling factors (sc), i.e.,
for different amplitudes of the torsional oscillation. }\label{proxy_nature}
\end{figure}

\begin{figure}
  \centering
  \includegraphics[width=8 cm,height=6 cm]{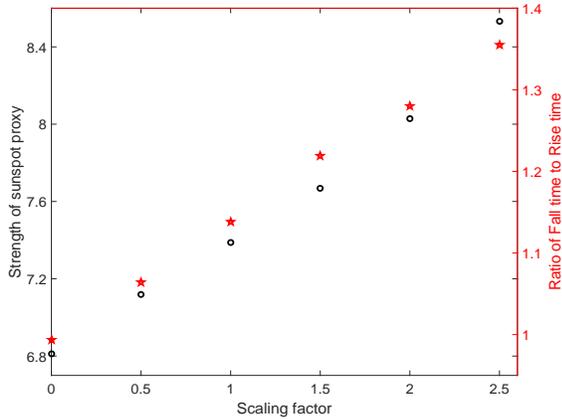}
  \caption{\textit{Black (circles):} The variation of peak value of the square of toroidal field (in
$Gauss^{2}$) during each cycle at the base of the convection zone
($r=0.7R_{\odot}$) at 15.3 degrees latitude with the scaling factor(amplitude) of
torsional oscillations. The peak of the square of toroidal field (sunspot proxy)
increases monotonically with the scaling factor.\newline
\textit{Red (stars):}  The variation of the ratio of fall time to rise time of sunspot proxy
for different scaling factors.}\label{sc_vs_amp}
\end{figure}

\begin{figure}
  \centering
  \includegraphics[width=8 cm]{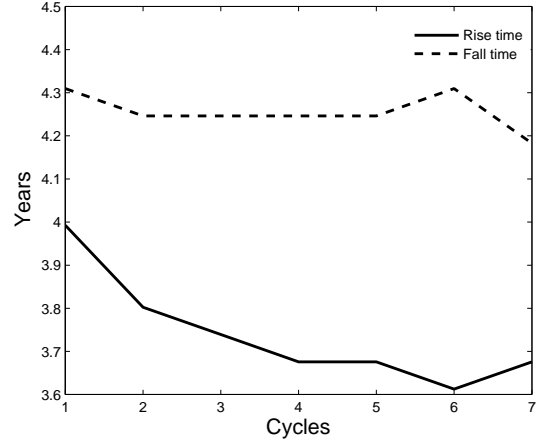}
  \caption{The variation of rise time and the fall time of sunspot proxy for a
scaling factor of unity for seven sunspot cycles in time. The output of the flux
transport dynamo shows some fluctuations, but still the rise time is always
significantly shorter than the fall time.}\label{tor1_risefall}
\end{figure}

\begin{figure}
  \centering
  \includegraphics[width=8 cm]{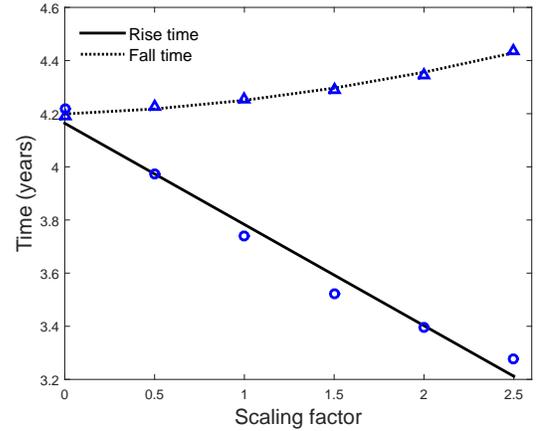}
  \caption{The rise time and fall time averaged over seven sunspot cycles for different scaling factors. The rise time of sunspot proxy monotonically decreases with increasing scaling factor while the fall time increases.}\label{risefalltime}
\end{figure}

 \hspace{5mm} \citet{2000ApJ...543.1027C} have proposed that the magnetic energy
density at about $15^{o}$ latitude at the base of the convection zone
($r=0.7R_{\odot}$) serves as a good proxy for the sunspot number. Since the
magnetic energy density is proportional to $B^{2}$ (where B is the toroidal magnetic field), we have
considered the peak value of $B^{2}$ as a proxy for the strength of
the sunspot cycle and explored how this changes with varying amplitude of
torsional oscillations. Henceforth, we use the term sunspot proxy to indicate
the
value of $B^{2}$ at 15.3 degrees latitude at the base of the convection zone
($r=0.7 R_{\odot}$).

The temporal shape of the solar cycle for different amplitudes of torsional oscillations are shown
in Fig. \ref{proxy_nature}. The variation of the peak value of sunspot proxy is
shown in Fig. \ref{sc_vs_amp} for different scaling factors. The monotonic
increase shown by the peak of the sunspot proxy on increasing the amplitude of
torsional oscillations indicates that these oscillations enhance the mechanism
for the production of toroidal field by increasing the shear in azimuthal
rotation.

\subsection{Impact of the amplitude of torsional oscillations on the rise and fall times of the solar cycle}\label{sectionrisefall}


Apart from increasing the peak value of the sunspot proxy, increase
in the amplitude of torsional oscillations also changes the nature of the
sunspot proxy. To learn more about this nature, we would like to define two
quantities here:
\vspace{2mm}
\par
1. Rise time: Rise time is defined as the time difference between the phases of
the cycle when the sunspot proxy increases from zero to 85\% of its maximum
strength.
\vspace{2mm}
\par
2. Fall time: Fall time is defined as the time difference between the phases of
the cycle when the sunspot proxy decreases from 85\% of its maximum strength to
zero.
\vspace{2mm}
\par
Without torsional oscillations the sunspot proxy is symmetric about its peak,
i.e., it has almost equal rise time (4.22 years) and fall time (4.19 years).
When torsional oscillations are introduced, the rise time becomes significantly
shorter than the fall time. Although the solution for the toroidal field from
the
model shows some fluctuations after introducing torsional oscillations, the rise
time is still always shorter than the fall time for all scaling factors. The
rise and fall times for sunspot cycles for a scaling factor of unity are
shown in Fig. \ref{tor1_risefall}. To check for robustness, we run the simulations
with all scaling factors for seven sunspot cycles. We find that rise time is
always
shorter than fall time for all scaling factors and in every cycle. The variation
of rise time averaged over seven sunspot cycles with the scaling factor is shown
in
Fig. \ref{risefalltime}. As we increase the amplitude of torsional oscillations,
the
rise time becomes shorter and shorter, while the fall time increases (because
the total period is fixed by the meridional flow). The ratio of fall time to
rise time is thus always greater than unity and it increases with increasing
amplitude of
torsional oscillations as shown in Fig. \ref{sc_vs_amp}. Thus, the greater the
amplitude of torsional oscillations, greater is the asymmetry in rise and fall
times of the sunspot cycle.

\subsection{The Waldmeier effect}\label{wald}


In 1935, \citet{17313} noted the following intriguing characteristics of the
sunspot cycle from observations of sunspot numbers which later came to be known as the Waldmeier effect.
\vspace{1mm}
\par
 1. The rise time of an average solar cycle is smaller than its fall time.
 \vspace{1mm}
\par
 2. The strength of a cycle is anticorrelated to its rise time. Shorter the rise
time, stronger is the cycle.
 \vspace{1mm}
\par
 3. The strength of a cycle is correlated to its rise rate. The higher the rise
rate, the stronger is the cycle.

\begin{figure}
  \centering
  \includegraphics[width=8 cm]{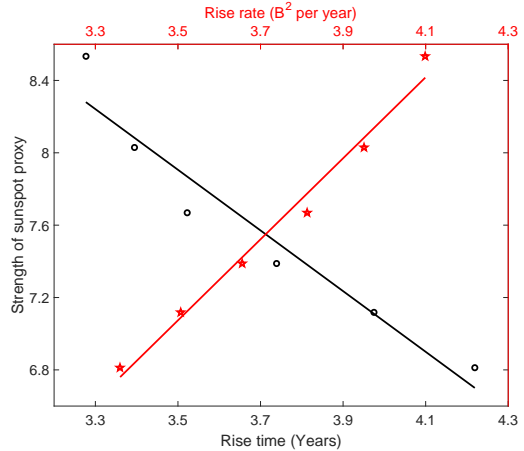}\\
  \caption{\textit{Black (circles):} The strength of the sunspot proxy ($Gauss^{2}$) versus its rise
time. This figure shows the classic signature of Waldmeier effect, which states
that sunspot cycles with shorter rise times tend to have greater
strength. \textit{Red (stars):} The strength of the sunspot proxy ($Gauss^{2}$) versus its rise
rate ($Gauss^{2}year^{-1}$). This figure shows the signature of the third
characteristic of Waldmeier effect, which states that sunspot cycles with higher
rise rates tend to have greater strength.}\label{waldmeier}
\end{figure}

\vspace{1mm}
\par
  These characteristics are together called the ``Waldmeier effect''.
  We have already seen the first characteristic in section
\ref{sectionrisefall}. In sections \ref{sectionamp} and \ref{sectionrisefall},
we have seen that the rise time of a sunspot cycle decreases while its strength
increases with increase in the amplitude of torsional oscillations. Thus, if we
plot the strength of the sunspot proxy versus its rise time, we can clearly see
the second characteristic of the Waldmeier effect (Fig. \ref{waldmeier}).

\par
  To check if we can see the third characteristic of Waldmeier effect, we adopt the following definition:\newline
\par
  Rise rate: The ratio of 85\% of the peak of the sunspot proxy to its rise
time.\newline

\par
  We know the rise time and the peak of the sunspot proxy from Figs.
\ref{risefalltime} \& \ref{sc_vs_amp}. Thus, we can calculate the rise rate
from
the above definition and plot it against the strength of the cycle (Fig.
\ref{waldmeier}). The red plot in Fig. \ref{waldmeier} is a clear confirmation of the
third characteristic of the Waldmeier effect. It shows that cycles with higher
rise rate tend to have greater strength.

\subsection{Theoretical Explanation: How do torsional oscillations contribute to the Waldmeier effect?}\label{explanation}

The rise time of solar cycles depends directly on the magnetic induction time scale where as their fall time depends mainly on diffusion or flux cancellation time scale. From Fig. \ref{risefalltime} it is clear that torsional oscillations impact the rise time much more than the fall time indicating that torsional oscillations enhance the magnetic induction process early in the solar cycle. To analyze their effect on the induction process, we simplify the toroidal field evolution equation (Eq. \ref{tor}) to focus on
the impact of torsional oscillations and recast it in the form

\begin{equation}
\frac{\partial B_{\phi}}{\partial t}= s(B_{p}.\nabla)\Omega + \chi
\end{equation}

where $\chi=$ contribution from all remaining terms (which do not directly
depend on the rotation and are responsible for advection and diffusion of the magnetic field). Expanding the first term related to the impact of
rotational shear on the toroidal field induction we get

\begin{equation}
 \frac{\partial B_{\phi}}{\partial t}= r\sin\theta\bigg(B_{r}\frac{\partial
\Omega}{\partial r}+\frac{B_{\theta}}{r}\frac{\partial \Omega}{\partial
\theta}\bigg)+\chi
\end{equation}

\begin{equation}
  \frac{\partial B_{\phi}}{\partial t}= B_{r}\bigg(r\sin\theta\frac{\partial\Omega}{\partial
r}\bigg)+ B_{\theta}\bigg(\sin\theta\frac{\partial\Omega}{\partial\theta}\bigg)+\chi
\end{equation}

Utilizing Eq. \ref{omega},

\begin{eqnarray}
 \frac{\partial B_{\phi}}{\partial t}=B_{r}\bigg(r\sin\theta\frac{\partial\Omega_{o}}{\partial r}\bigg)+
B_{\theta}\bigg(\sin\theta\frac{\partial\Omega_{o}}{\partial\theta}\bigg)+ \nonumber \\
 B_{r}\bigg(r\sin\theta\frac{\partial(\delta\Omega)}{\partial r}\bigg)+
B_{\theta}\bigg(\sin\theta\frac{\partial(\delta\Omega)}{\partial\theta}\bigg)+\chi
\end{eqnarray}

Thus, the difference in growth rate of the toroidal magnetic field (B) between a
solar cycle with torsional oscillations and one without torsional oscillations
(denoted by subscript ``o'') will be:

\begin{eqnarray}
 \bigg(\frac{\partial B_{\phi}}{\partial t}\bigg)-\bigg(\frac{\partial B_{\phi}}{\partial
t}\bigg)_{o}\approx B_{r}\bigg(r\sin\theta\frac{\partial(\delta\Omega)}{\partial r}\bigg)
\nonumber \\
+B_{\theta}\bigg(\sin\theta\frac{\partial(\delta\Omega)}{\partial\theta}\bigg)
\end{eqnarray}

\begin{figure}
  \centering
  \includegraphics[width=8 cm]{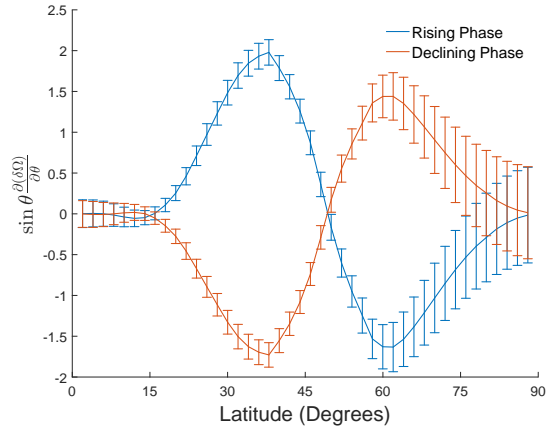}\\
  \caption{The colatitudinal gradient of torsional oscillations at the base of
the solar convection zone ($0.7R_{\odot}$)
  averaged over the rising phase (1995.5 to 2001) and over the declining phase
(2001 to 2007.2). Torsional oscillations increase
  the shear in rotation during the rising phase and decrease the shear during
the declining phase at latitudes lower than 50 degrees.
The increase in shear during the rising phase leads to a stronger and more
steeply rising solar cycle.}\label{lat_gradDR}
\end{figure}

Most kinematic flux transport dynamo models generate toroidal magnetic field primarily from $B_{r}$ at high latitudes (above $60$  degrees) and from $B_{\theta}$ (where $\theta$ is the colatitude) at lower latitudes between $0.7R_{\odot}$ and $0.8R_{\odot}$ inside the Sun\citep{2009ApJ...698..461M}. Since we are considering only the $\Omega$-effect here on the sunspot proxy around $15$ degrees latitude and at the bottom of the convection zone, $B_{\theta}$ dominates over $B_{r}$. If one starts with positive polar flux at solar minimum near the north pole, this implies a positive $B_{r}$ but a negative $B_{\theta}$ in the northern hemisphere. At low latitudes near the base of the convection zone the sign of the co-latitudinal derivative of $\Omega$ is positive resulting in the production of a negative $B_{\phi}$ which is the our proxy for sunspots. The average observed magnitude of $\frac{d(\delta\Omega)}{d\theta}$ from helioseismic measurements is larger during the rising phase of the cycle compared to the declining phase (see Fig. \ref{lat_gradDR}) at low latitudes and $0.7R_{\odot}$. Thus, during the rising phase over a significant region of the convection zone, the growth rate of the magnetic field is higher
with torsional oscillations than without -- which explains both the steeper rise and the increasing strength seen
in solar cycles with higher amplitude of torsional
oscillations. This direct relationship between
the strength of torsional oscillations and the growth rate (and strength) of the
toroidal magnetic field -- contributes to the Waldmeier effect. This explains why we
recover all the characteristics of the Waldmeier effect in our dynamo simulations
driven by the observed torsional oscillations without recourse to any other physics.

It should be noted here that the rise-rate of the cycle depends on the product of the strength of the polar field from the previous cycle ($B_{r}$, $B_{\theta}$) and the gradient of $\Omega$. Since we do not see a dramatic change (more than a few percent) in the strength of the solar cycle upon the inclusion of torsional oscillations, most of the modulation in strength of the solar cycle may be arising from modulation of the strength of polar field at the end of the previous cycle. Compelling evidence for such a correlation between the polar field of the previous cycle and the strength of the solar cycle has been presented by various authors \citep{1978GeoRL...5..411S,2012ApJ...761L..13K,2013ApJ...767L..25M}. Inclusion of torsional oscillations thus primarily makes the sunspot cycles rise faster, asymmetric around their peak and contributes less towards modulation of their amplitude.


\section{Concluding Discussions}\label{conclusion}

Previous attempts of simulating the Waldmeier effect with flux transport dynamos
have relied on stochastic fluctuations in the dynamo parameters (poloiodal
source and meridional circulation)\citep{2011MNRAS.410.1503K,2013IAUS..294..595P}; see also
\citet{2000ApJ...543.1027C}. While these fluctuations can certainly contribute to the Waldmeier effect, we have shown that the contribution from torsional oscillations needs to be taken into account as well. We have successfully reproduced all the three
characteristics of Waldmeier effect qualitatively by incorporating the observed torsional oscillations and modulating their amplitude without recourse to any changes in other dynamo parameters, including cycle period. The analytical explanation based on induction equation convincingly establishes the theoretical basis of how cycle to cycle variations in torsional oscillations can contribute to the observed Waldmeier effect. Using torsional oscillation data for solar cycle 23 we get a fall time to rise time ratio of 1.14 in our simulations which is less than the observed ratio of 1.7 recorded for solar cycle 23. We note that this does not necessarily mean torsional oscillations contribute less to the Waldmeier effect than other processes because making a quantitative comparison requires a model calibrated with observations. One would require the correct turbulent diffusivity profile, meridional flow profile and measurements of $B_{r}$ and $B_{\theta}$ in the solar interior to perform a rigorous quantitative analysis of the extent of the contribution of torsional oscillations to the Waldmeier effect. We emphasize that our theoretical explanation is based on the magnetic induction equation and thus independent of any specific dynamo model.

We also conducted some additional numerical simulations to investigate which branch of torsional oscillations has the dominant effect on the shape of the solar cycle. In one set of simulations, we turned off torsional oscillations in the deep interior (below $0.85R_{\odot}$) and in another set we turned off the torsional oscillations in the upper half of the convection zone (above 0.85$R_{\odot}$). The simulations with torsional oscillations only in the upper half of the convection zone did not reproduce the Waldmeier effect and it looked like the torsional oscillations did not alter the solar cycle in any fashion. The set of simulations with torsional oscillations only in the bottom half of the convection zone on the other hand produced results that are identical to the simulations with torsional oscillations introduced in the entirety of the convection zone. This makes sense because the location of magnetic field amplification in these simulations coincides with the deeper branch of torsional oscillations. Another experiment was conducted to independently investigate the effect of the high latitude and low latitude branches of torsional oscillations on the solar cycle. The division between these branches was made at $60^{o}$ latitude. The low latitude branch seems to be the one which has the dominating effect on the shape of the solar cycle. It produced a fall time to rise time ratio of $1.10$ where as the high latitude branch produced a ratio of $1.03$ compared to $1.14$ - the ratio obtained for torsional oscillations in the entire convection zone. This indicates that magnetic field - plasma flow interactions in the low-latitude and deeper layers of the convection zone contribute more towards the Waldmeier effect.

Our results show that increasing amplitude of torsional oscillations increase
the strength of
sunspot cycles (Fig. \ref{sc_vs_amp}) as well as their rise rate and decrease their rise times (Fig.
\ref{risefalltime}). Thus, the amplitude of torsional oscillations may act as the
connection between the strength of the cycle and its rise time noted by
\citet{17313} (Fig. \ref{waldmeier}).

These results have important implications for
solar cycle predictions. A reasonably successful precursor technique for solar
cycle predictions is based on observations of the early growth-rate of the cycle
in question \citep{2008ApJ...685.1291C,2008SoPh..252..209P}. There
is also an independent, and not yet rigorously proven, understanding emerging
within the scientific community that the nature of torsional oscillation
patterns of the extended solar cycle \citep{2014ApJ...792...12M} can indicate the strength of upcoming cycles \citep{2009ApJ...701L..87H,2015TESS....110502H}. Our theoretical analysis and dynamo simulations
causally connect torsional oscillations to the growth rate and amplitude of
sunspot cycles, thereby providing a physical basis for solar cycle predictions
based on the Waldmeier effect or early observations of torsional oscillations of the extended solar cycle.

\acknowledgements{Computational facilities at the Center of Excellence in Space
Science India (CESSI) were utilized for the purposes of this research. CESSI is
funded by the Ministry of Human Resource Development, Government of India. This
research was supported in parts through a NASA Grand Challenge grant NNX14AO83G and the Indo-U.S. Joint Center Programme grant IUSSTF-JC-011-2016. This work utilizes data obtained by the Global Oscillation Network Group (GONG) project, managed by the National Solar Observatory, which is supported by AURA, Inc., under a cooperative agreement with the National Science Foundation. The data were acquired by instruments operated by the Big Bear Solar Observatory, High Altitude Observatory, Learmonth Solar Observatory, Udaipur Solar Observatory, Instituto de Astrofisico de Canarias and Cerro Tololo Inter-American Observatory. The authors would also like to thank Dr. Matthias Rempel for his helpful comments which improved this paper.}

\bibliographystyle{apj}
\bibliography{references}



\end{document}